# *DECO2* – An Open-source Energy System Decarbonisation Planning Software Including Negative Emissions Technologies

Purusothmn Nair S. Bhasker Nair[1], Raymond R. Tan[2], Dominic C. Y. Foo[1], Disni Gamaralalage[3], Michael Short[4] *

[1] Department Chemical and Environmental Engineering/Centre of Excellence for Green Technologies, University of Nottingham Malaysia, Broga Road, 43500 Semenyih, Selangor, Malaysia

[2] Department of Chemical Engineering, De La Salle University, 2401 Taft Avenue, 0922 Manila, Philippines

[3] Presidential Endowed Chair for "Platinum Society", The University of Tokyo, 7-3-1 Hongo, Bunkyo-ku, Tokyo 113-8656, Japan

[4] Department of Chemical and Process Engineering, University of Surrey, Guildford, Surrey GU2 7XH, United Kingdom

Abstract

The deployment of $CO_2$ capture and storage (CCS) and negative emissions technologies (NETs) are crucial to meet the net-zero target by year 2050, as emphasised by the Glasgow Climate Pact. Over the years, several energy planning models have been developed to address the temporal aspects of carbon management. However, limited works have incorporated CCS and NETs for bottom-up energy planning at the individual plant scale, which is considered in this work. The novel formulation is implemented in an open-source energy system software that has been developed in this work for optimal decarbonisation planning. The DECarbonation Options Optimisation (*DECO2*) software considers multiperiod energy planning with a superstructural model and was developed in Python with an integrated user interface in Microsoft Excel. The software application is demonstrated with two scenarios that differ in terms of the availabilities of mitigation technologies. Results demonstrated the potential of fuel substitutions for low-carbon alternatives in existing coal and natural gas power plants. Additionally, once NETs are mature and are available for commercial deployment, their deployment is crucial in aiding $CO_2$ removal in minimal investment costs scenarios. Overall, the newly developed open-source software demonstrates the importance of determining the optimal deployment of mitigation technologies in meeting climate change targets for each period.

## 1. Introduction

The Sustainable Development Goals of the United Nations consist of 17 goals meant for the improvement of the planet, as well as the enhancement in the quality of people's life (United Nations, 2021a). The goals under Sustainable Development Goals are often interlinked. For example, the reduction in global poverty should be aligned with the enhancement of the health, education and economic sectors (United Nations, 2021a). Of the 17 goals, Goal 13 is related to climate actions, which demands urgent actions to mitigate

*Corresponding author. Emails: m.short@surrey.ac.uk (Short M), dominic.foo@nottingham.edu.my (DCYF), purusothmn.nair@gmail.com (PNSBN), raymond.tan@dlsu.edu.ph (RR Tan), disnigm@gmail.com (Gamaralalage D)

climate change impacts (United Nations, 2021a). In November 2015, 196 countries joined forces for the adoption of an international treaty on climate change (United Nations Climate Change, 2021a). The Paris Agreement aimed to limit global warming to within 2 °C and preferably 1.5 °C above pre-industrial levels (United Nations Climate Change, 2021a). Despite this agreement, countries have shown limited signs of reducing greenhouse gas (GHG) emissions. As of 2020, the average surface temperature of Earth was 1.2 °C higher in comparison to the pre-industrial period (United Nations, 2021b). As the $CO_2$ concentration in the atmosphere exceeded 410 ppm in 2020, emissions will continue to rise unless drastic mitigation actions are taken (United Nations, 2021b). Despite the fall in $CO_2$ emissions during the COVID-19 pandemic, this trend was only short-lived (United Nations, 2021b). By December 2020, the GHG concentrations were 2% higher than the level recorded 12 months ago (United Nations, 2021b). At the current trajectory, a shift in the economy towards carbon neutrality must occur to prevent a further rise in GHG emissions (United Nations, 2021b). According to the IPCC (2022a), carbon neutrality can only be realistically achieved with the use of *negative emissions technologies* (NETs) in addition to other climate change mitigation measures. Any delay in mitigation actions would only compound the impacts of climate change e.g., rising sea levels, prolonged droughts, greater precipitation etc.

The critical impacts of climate change necessitate efficient energy planning modelling tools. On that note, this work develops an open-source energy system software that can be used for optimal decarbonisation planning. The mathematical formulation in this work was inspired by *carbon emissions pinch analysis* (CEPA) that was developed by Tan and Foo (2007) to determine the minimum deployment of renewable energy sources in meeting the $CO_2$ emissions limit of a geographical region. Later, Tan et al. (2009) extended the graphical targeting technique to incorporate $CO_2$ capture and storage (CCS). However, these earlier works were limited by being unable to consider time factors. Practically, energy planning should be conducted over a broad time horizon, which may take several years or even decades. Such energy planning models should be able to handle the temporal aspects of carbon management, involving variations in demand and $CO_2$ emission limits. On that note, a multiperiod energy planning model (Ooi et al., 2014) was developed with the use of the automated targeting method (ATM) (Ooi et al., 2013). In comparison to the graphical technique, mathematical programming approaches are often preferable in the development of energy planning models due to their abilities in handling large-scale problems. Earlier work on the superstructural model was reported by (Pękala et al., 2010) for the deployment of biofuels and CCS in the transportation and power generation sectors. Later, a fuzzy integer programming model was formulated to account for the environmental and economic constraints during CCS deployment (Tan et al., 2010). Subsequently, the optimal matching of $CO_2$ sources to available sinks was done via a continuous-time mixed-integer non-linear programming (MINLP) model (Tan et al., 2012). In this work, the MINLP model was simplified to a mixed-integer linear programming (MILP) model by assuming a fixed flow rate of $CO_2$ source with no constraints imposed on the $CO_2$ storage capacities (Tan et al., 2012). The practicality of this work was later enhanced by considering varying $CO_2$ flow rates and the existence of a limit on the $CO_2$ storage capacities (Tan et al., 2013). The work by Tan et al. (2013) was also made up of a multiperiod model for realistic energy planning involving CCS deployment. Following this, a further extension to the multiperiod MILP model was done by considering unequal time intervals (Lee et al., 2014).

Despite the deployment of both renewable energy sources and CCS, it would still require the use of $CO_2$ removal (CDR) via NETs for limiting global warming (IPCC, 2022). The lack of drastic actions in previous years means that CDR is necessary for limiting warming to preferably within 1.5 °C above pre-industrial levels. CDR may occur via two types of NETs i.e., energy-producing NETs (EP-NETs) and energy-consuming NETs (EC-NETs). The former generates energy during $CO_2$ load removal. Some examples of EP-NETs are bioenergy with CCS (BECCS) and biochar (McGlashan et al., 2012). By contrast, there is an energy penalty associated with $CO_2$ load removal for the latter technology. Direct air capture (DAC) (McGlashan et al., 2012), enhanced weathering (EW) (Seifritz, 1990) and ocean liming (McLaren, 2012) are some of the technologies associated with EC-NETs. In the past, mathematical programming and pinch-based approaches had been developed to consider the deployment of NETs during carbon-constrained energy planning. For the latter approach, the graphical targeting technique (Tan et al., 2009) was extended to incorporate EP-NETs during energy planning (Nair et al., 2020). Recognising that a portfolio of NETs would be required to accelerate CDR, this work was later revamped for the combined deployment of both EP-NETs and EC-NETs (Nair et al., 2022). The limitations of the graphical targeting technique were overcome with the development of an algebraic targeting technique where renewable energy sources, CCS and NETs were considered during energy planning (Nair et al., 2021). In terms of mathematical programming approaches, an optimal source-sink matching for CDR via biochar was initially conducted via a MILP model (Tan, 2016). The objective function of the model was set for the maximisation of CDR without compromising the soil quality (Tan, 2016). Other work involving biochar was done via a fuzzy linear programming model involving biomass co-firing in power plants (Aviso et al., 2020). Aside from biochar, EW is also an effective means of CDR, with several mathematical programming approaches being developed previously to account for its deployment (Seifritz, 1990). The optimisation of EW networks initially took place via a linear programming model (Tan and Aviso, 2019). Due to the uncertainties with EW networks in terms of silicate rock grinding and property variations, a fuzzy MINLP model was formulated to address these issues (Aviso and Tan, 2020). The fuzzy model was further enhanced to consider the uncertainties that exist within industrial supply chains and economic evaluations (Aviso et al., 2021). A recent work considered the use of non-hazardous industrial waste during EW, where a superstructural model was developed for its evaluation and analysis (Jia et al., 2022). Once again, recognising the $CO_2$ drawdown would require a portfolio of NETs e.g., BECCS, DAC, EW and biochar, a linear programming model was formulated to optimise NETs deployment under resource constraints e.g., land, water, nutrients and energy (Migo-Sumagang et al., 2021). Due to a projected extensive deployment of EW, a stand-alone supply chain like system would exist, thus presenting a need for its optimisation. Therefore, a more recent work aimed to use a MILP model for the optimisation of the processes that occur in a EW network (Tan et al., 2022).

Given the urgent decarbonisation initiatives, energy planning tools have been developed to aid in policymaking and the planning for future energy generation. Although the pinch analysis approach is beneficial in terms of setting high-level targets (e.g., total deployment of renewable energy sources), analytical tools are more equipped to conduct detailed energy planning (e.g., the type of plant that would require CCS installation). On that note, the two major energy planning modelling tools that are typically employed are the bottom-up and top-down models. The former model is focused on the components of an energy planning system i.e., the availability of various technologies, and the overall costs involved (Prina et al., 2020). In other words, the target demand and emission limits are satisfied based on the technologies made available in a period and the allocated budget (Prina et al., 2020). Meanwhile, the latter model investigates the impact of set demand and emissions limit targets on the economic and energy sectors (Indra Al Irsyad et al., 2017). In other words, the top-down model investigates economic impacts due to the implementation of such energy policies. In this work, a novel bottom-up model is developed due to the availability of a wide technology library, aside from cost constraints. Therefore, the discussion of existing energy planning models will be tailored to those that had employed the bottom-up model.

The first example of a bottom-up energy planning model is the Wien Automated System Planning (WASP) model (Jenkin and Joy, 1974). WASP is one of the first energy planning models developed for the expansion of power generation systems. A user can add energy planning constraints such as fuel availabilities, the requirement of system reliability and emissions limits (Jenkin and Joy, 1974). Based on these constraints, the model determines the optimal configurations for the expansion of existing energy systems (Jenkin and Joy, 1974). Note that the WASP model considers the costs involved with existing and new energy generation plants (Jenkin and Joy, 1974). Another available energy planning model is OSeMOSYS (Howells et al., 2011). Being an open-source energy modelling system, OSeMOSYS provides an integrated assessment for energy planning. Aside from providing detailed power configurations, this energy planning model also considers multi-resource systems i.e., economic, material and energy (Howells et al., 2011). More recently, OSeMOSYS integrated smart grids to deal with intermittent renewable energy sources (Welsch et al., 2012). One of the most common energy planning models in recent times is MARKAL. MARKAL is widely used in various countries for satisfying emissions constraints (IEA-ETSAP, 2022a). This energy planning model consists of a pool of technologies with their associated costs (IEA-ETSAP, 2022a). Based on the demand of the energy planning system, a range of technologies that minimise the total costs is selected (IEA-ETSAP, 2022a). Most importantly, this model does not require a definition of the ranking of GHG, as observed in other bottom-up energy planning models (IEA-ETSAP, 2022a). TIMES is another energy planning model which serves as an extension of MARKAL. Although the energy planning principles of TIMES are very much similar to MARKAL, the former model now integrates an economic approach to supplement the existing technical approach in MARKAL (IEA-ETSAP, 2022b). The solutions obtained from TIMES are based on scenario analysis. Initially, a base case scenario is developed. This reference scenario does not contain possible energy planning constraints e.g., minimum deployment of renewable energy, permissible emissions limits etc. (IEA-ETSAP, 2022b). Once the constraints are inputted, a separate scenario is then developed and compared against the base case scenario (IEA-ETSAP, 2022b). The benefits of TIMES are the possibility of viewing multiple scenarios which employ different sets of technologies, with varying costs (IEA-ETSAP, 2022b). The user could select the best possible energy planning scenario to be employed, based on one's preference. Table 1 presents the summary of the existing bottom-up energy planning models available for deployment.

From Table 1, it can be observed that in all existing energy planning models, little focus has been placed on the key role that NETs and CCS can play in the future. Note however that both of these technologies must be incorporated to align with the targets set during the Paris Agreement (United Nations Climate Change, 2021a) and the Glasgow Climate Pact (United Nations Climate Change, 2021b). Therefore, this work reports the development of a Decarbonisation Options Optimisation (*DECO2*) software which consists of a pool of technologies including CCS and NETs that may be employed to meet the $CO_2$ emissions limits at specific facilities, along with new generation capabilities and decommissioning strategies. The multiperiod energy planning model in this software may be employed by policymakers to determine long-term decarbonisation strategy. The latter includes the timeline for decommissioning of plants, technology implementation, and fuel substitutions for fossil-based power plants (to renewable energy sources). The mathematical programming in the software is expected to provide rigorous optimal solutions, subject to constraints such as the availability of low-carbon fuels, technology readiness, availability of renewable energy sources, etc. The paper is organised as follows. A formal problem statement is presented in the next section. This is then followed by the mathematical formulations for carbon-constrained energy planning. The software infrastructure is then presented to demonstrate the software application. Two scenarios of a hypothetical case study which differ in terms of the deployment of mitigation strategies are presented to show the applicability of the *DECO2* software. Finally, conclusions and prospects for future work are provided.

*Table 1: Summary of bottom-up energy planning models*

| ENERGY PLANNING MODELS | FEATURES | MISSING KEY FEATURE |
|---|---|---|
| WASP | - Power generation systems<br>- Fuel availabilities and emissions limits<br>- Optimal expansion of existing energy systems<br>- Cost of existing and new energy generation plants | CCS and NETs for individual plants |
| OSEMOSYS | - Integrated assessment for energy planning<br>- Multi-resource systems (economic, material & energy) | |
| MARKAL | - A pool of technologies for satisfying emissions and cost constraints | - NETs deployment<br>- Open-source software<br>- Commissioning and decommissioning strategies<br>- Technology implementation time<br>- Easy-to-use input spreadsheet |
| TIMES | - Extension to MARKAL<br>- Incorporates economic approach<br>- Viewing multiple energy planning scenarios | |

## 2. Problem Statement

The formal problem statement for the development of a process-integration-based software for optimal decarbonisation is as follows:

- A pool of fossil-based (coal and natural gas) and renewable-energy-based (solar, hydropower etc.) plants are available to satisfy the energy demand and $CO_2$ emissions limit of an energy planning system in period $k \in K$.
- Power plant *i* has a lower bound energy output ($F_{i,LB}$) upper bound energy output ($F_{i,UB}$) and $CO_2$ emissions intensity ($CS_i$) that make up the energy planning system for period $k \in K$.
- The commission and decommissioning periods for power plant *i* are specified.
- The total energy demand ($D_k$) and $CO_2$ emission limit ($L_k$) of the energy planning system in period $k \in K$ are specified.
- The removal of the $CO_2$ emissions in period $k \in K$ is aided by the deployment of renewable energy source $r \in R$, CCS technology $n \in N$, EP-NETs technology $p \in P$, EC-NETs technology $q \in Q$, alternative solid-based fuel $s \in S$ and alternative gas-based fuel $g \in G$.
- The main task is to determine the energy generation from power plant *i* and the minimum deployment of each technology (renewable energy, CCS, NETs and alternative fuel) in satisfying the demand and $CO_2$ emissions constraints of an energy planning system in period $k \in K$.

Figure 1 presents a superstructure representation of the process-integration-based software for optimal decarbonisation.

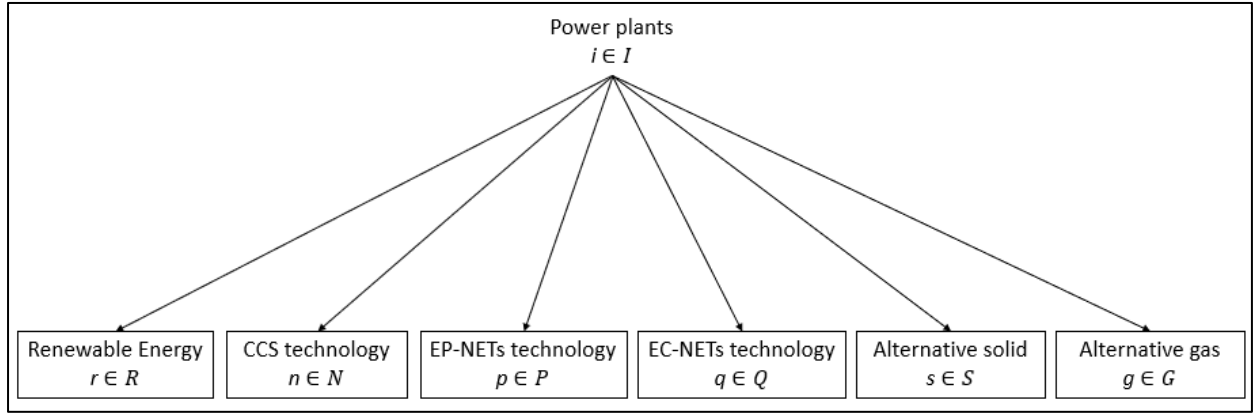

*Figure 1: Superstructure representation of the process-integration-based software for optimal decarbonisation*

## 3. Mathematical Formulation of *DECO2*

*DECO2* is based on a superstructural model, consisting of existing and upcoming power plants available for power generation, as well as a pool of mitigation technologies. The superstructural model initially determines the energy generation from power plant *i* that satisfies the power demand of energy planning period *k*. Following this, the optimal deployment of mitigation technologies i.e., renewable energy sources, CCS, NETs, and alternative fuels in period *k* is determined based on the demand and $CO_2$ emissions constraints. The superstructural model developed in this work would act as a guide to policymakers in terms of power plants' commissioning and decommissioning timelines and total costs involved in meeting the energy demand and $CO_2$ emissions limits of a geographical region.

Firstly, the cumulative deployment of energy source from power plant *i* ε *I* should satisfy the energy demand of period *k*; shown in Equation 1. Note that energy generation by power plant *i* in period *k* ($FS_{i,k}$) is constrained by its lower ($F_{i,LB}$) and upper bound of energy generation ($F_{i,UB}$) as demonstrated in Equation 2 and Equation 3 respectively.

$$\sum_i FS_{i,k} = D_k \qquad \forall k \qquad \text{Equation 1}$$

$$FS_{i,k} \geq F_{i,LB} \times A_{i,k} \qquad \forall i\ \forall k \qquad \text{Equation 2}$$

$$FS_{i,k} \leq F_{i,UB} \times A_{i,k} \qquad \forall i\ \forall k \qquad \text{Equation 3}$$

where $A_{i,k}$ is the binary variable for energy generation by power plant *i* in period *k*.

Next, the energy generation from power plant *i* in period *k* is subject to either its commissioning or decommissioning timeline as demonstrated in Equation 4. Energy generation from power plant *i* would only take place from the commissioning period ($CM_i$) onwards. Before the commissioning period, there should not be any energy generation from power plant *i*. By contrast, energy generation from power plant *i* would only take place until the period before the decommissioning period (*DCM$_i$*).

$$FS_{i,k} = \begin{cases} 0, & k < CM_i \\ 0, & k \geq DCM_i \end{cases} \quad \forall i\ \forall k \qquad \text{Equation 4}$$

Also, the energy generated from power plant *i* in period *k* should at least match its generation in the previous period, as shown in Equation 5. This constraint ensures a continuous operation of power plant *i* if it is selected for power generation. A temporary shutdown of power plant *i* would be impractical, thus resulting in a negative return on investment.

$$FS_{i,k+1} \geq FS_{i,k} \quad \forall i\ ;\ k = 1,2,\ldots,n-1 \qquad \text{Equation 5}$$

CCS is one of the mitigation technologies that may be employed in power plant *i* for the satisfaction of the $CO_2$ emissions limits (IEA, 2021). CDR via CCS decreases the $CO_2$ intensity of power plant *i* (*CS$_i$*). Therefore, the $CO_2$ intensity of power plant *i* with the deployment of CCS technology *n* in period *k* (*CR$_{i,k,n}$*) is calculated from Equation 6.

$$CR_{i,k,n} = \frac{CS_i \times (1 - RR_{k,n})}{1 - X_{k,n}} \quad \forall i\ \forall k\ \forall n \qquad \text{Equation 6}$$

where *RR$_{k,n}$* and *X$_{k,n}$* represent the removal ratio and parasitic power loss of CCS technology *n* in period *k* respectively.

The deployment of CCS technology *n* is constrained by the upper bound of energy generation by power plant *i*; shown in Equation 7. Also, the cumulative deployment of all CCS technologies should not exceed the energy generation by power plant *i* in period *k* as demonstrated in Equation 8.

$$FR_{i,k,n} \leq F_{i,UB} \times B_{i,k,n} \quad \forall i\ \forall k\ \forall n \qquad \text{Equation 7}$$

$$\sum_n FR_{i,k,n} \leq FS_{i,k} \quad \forall i\ \forall k \qquad \text{Equation 8}$$

where *FR$_{i,k,n}$* is the deployment of CCS technology *n* in power plant *i* in period *k* while *B$_{i,k,n}$* is the binary variable for the deployment of CCS technology *n* in power plant *i* in period *k*.

Like the energy generation from power plant *i*, the deployment of CCS technology *n* in power plant *i* in period *k* should at least match its deployment in the previous period as shown in Equation 9. CCS technology is capital-intensive. Therefore, it is not practical and economically unviable for CCS technology *n* to be deployed in period *k* and not used in the subsequent period.

$$FR_{i,k+1,n} \geq FR_{i,k,n} \quad \forall i\ \forall n\ ;\ k = 1,2,\ldots,n-1 \qquad \text{Equation 9}$$

CCS deployment incurs parasitic power loss of energy sources. Therefore, the net energy of power plant *i* after the deployment of CCS technology *n* in period *k* (*FNR$_{i,k,n}$*) is calculated from Equation 10.

$$FR_{i,k,n} \times (1 - X_{k,n}) = FNR_{i,k,n} \qquad \forall i \ \forall k \ \forall n \qquad \text{Equation 10}$$

Aside from CCS, mitigation technologies that are available for power plant *i* in this work are alternative solid-based fuel *s* and alternative gas-based fuel *g*. Note that alternative solid and gas-based fuels may be used to replace fuels in power plant *i* which are in solid and gas phases respectively. The deployment of these alternative fuels in power plant *i* in period *k* should at least match its deployment in the previous period as shown in Equation 11 and Equation 12. The reasoning for this is the same as for CCS deployment.

$$FAS_{i,k+1,s} \geq FAS_{i,k,s} \qquad \forall i \ \forall s \ ; \ k = 1,2,\ldots,n-1 \qquad \text{Equation 11}$$

$$FAG_{i,k+1,g} \geq FAG_{i,k,g} \qquad \forall i \ \forall g \ ; \ k = 1,2,\ldots,n-1 \qquad \text{Equation 12}$$

where $FAS_{i,k,s}$ and $FAG_{i,k,g}$ are the deployment of alternative solid-based fuel *s* and gas-based fuel *g* in power plant *i* in period *k* respectively.

Also, the deployment of alternative solid-based fuel *s* and gas-based fuel *g* in power plant *i* in period *k* is constrained by the upper bound of power generation by power plant *i*, as shown in Equation 13 and Equation 14 respectively. The use of alternative fuels should never exceed the maximum power generation by power plant *i* as these low $CO_2$ intensity fuels are only purposed to replace the higher $CO_2$ intensity fuels that were originally deployed.

$$FAS_{i,k,s} \leq F_{i,UB} \times G_{i,k,s} \qquad \forall i \ \forall k \ \forall s \qquad \text{Equation 13}$$

$$FAG_{i,k,g} \leq F_{i,UB} \times H_{i,k,g} \qquad \forall i \ \forall k \ \forall g \qquad \text{Equation 14}$$

where $G_{i,k,s}$ and $H_{i,k,g}$ are the binary variables for the deployment of alternative solid-based fuel *s* and gas-based fuel *g* in power plant *i* in period *k* respectively.

The cumulative deployment of all mitigation technologies available in this work should equate to the energy generated by power plant *i* in period *k* ($FS_{i,k}$) as demonstrated in Equation 15. The latter is initially determined from Equation 1.

$$FNS_{i,k} + \sum_n FR_{i,k,n} + \sum_s FAS_{i,k,s} + \sum_g FAG_{i,k,g} = FS_{i,k} \qquad \forall i \ \forall k \qquad \text{Equation 15}$$

where $FNS_{i,k}$ is the net energy of power plant *i* without the deployment of mitigation technologies.

Other mitigation technologies that are available in this work are renewable energy source *r*, EP-NETs technology *p* and EC-NETs technology *q*. Note that these technologies are not plant-specific. Instead, the cumulative deployment of these mitigation technologies is determined for period *k*. The deployment of renewable energy source *r* ($FC_{k,r}$), EP-NETs technology *p* ($FEP_{k,p}$) and EC-NETs technology *q* ($FEC_{k,q}$) in period *k* are constrained by the availability of each technology, as demonstrated in Equation 16, Equation 17 and Equation 18 respectively.

$$FC_{k,r} \leq AC_{k,r} \times C_{k,r} \quad \forall k \, \forall r \quad \text{Equation 16}$$

$$FEP_{k,p} \leq AEP_{k,p} \times D_{k,p} \quad \forall k \, \forall p \quad \text{Equation 17}$$

$$FEC_{k,q} \leq AEC_{k,q} \times E_{k,q} \quad \forall k \, \forall q \quad \text{Equation 18}$$

where $C_{k,r}$, $D_{k,p}$ and $E_{k,q}$ are the binary variables for the deployment of renewable energy source *r*, EP-NETs technology *p* and EC-NETs technology *q* in period *k* respectively and $AC_{k,r}$, $AEP_{k,p}$, $AEC_{k,q}$ are the availabilities of renewable energy source *r*, EP-NETs technology *p* and EC-NETs technology *q* in period *k* respectively.

Similar to the CCS and alternative fuels, the deployment of renewable energy source *r*, EP-NETs technology *p* and EC-NETs technology *q* in period *k* should at least match its deployment in the previous period; demonstrated in Equation 19, Equation 20 and Equation 21 respectively. Should a plant be commissioned in period *k*, it is economically viable for its operation to be continuous in subsequent periods to ensure a positive return on investment.

$$FC_{k+1,r} \geq FC_{k,r} \quad \forall r\,;\, k = 1,2,\ldots,n-1 \quad \text{Equation 19}$$

$$FEP_{k+1,p} \geq FEP_{k,p} \quad \forall p\,;\, k = 1,2,\ldots,n-1 \quad \text{Equation 20}$$

$$FEQ_{k+1,q} \geq FEQ_{k,q} \quad \forall q\,;\, k = 1,2,\ldots,n-1 \quad \text{Equation 21}$$

The cumulative deployment of all mitigation technologies (CCS, alternative fuels, renewable energy sources and NETs) should satisfy the total demand of the energy system of period *k*; the latter includes the total power requirement ($D_k$) and that required by EC-NETs ($FEC_{k,q}$), as demonstrated in Equation 22.

$$\sum_i FNS_{i,k} + \sum_i \sum_n FNR_{i,k,n} + \sum_i \sum_s FAS_{i,k,s} + \sum_i \sum_g FAG_{i,k,g} + \sum_r FC_{k,r} + \sum_p FEP_{k,p} = \sum_q FEC_{k,q} + D_k \quad \forall k \quad \text{Equation 22}$$

Following this, the total $CO_2$ load contribution from all power plants and mitigation technologies of energy planning period *k* ($TE_k$) is determined from Equation 23.

$$\sum_i FNS_{i,k}\,CS_i + \sum_i \sum_n FNR_{i,k,n}\,CR_{i,k,n} + \sum_i \sum_s FAS_{i,k,s}\,CIAS_{k,s} + \sum_i \sum_g FAG_{i,k,g}\,CIAG_{k,g} + \sum_r FC_{k,r}\,CIC_{k,r} + \sum_p FEP_{k,p}\,CIEP_{k,p} + \sum_q FEC_{k,q}\,CIEC_{k,q} = TE_k \quad \forall k \quad \text{Equation 23}$$

where $CIAS_{k,s}$, $CIAG_{k,g}$, $CIC_{k,r}$, $CIEP_{k,p}$ and $CIEC_{k,q}$ represent the $CO_2$ intensities of alternative solid-based fuel $s$, alternative gas-based fuel $g$, renewable energy source $r$, EP-NETs technology $p$ and EC-NETs technology $q$ in period $k$ respectively.

Next, the total costs of power generation by power plant $i$ in period $k$ ($CTF_k$) are calculated from Equation 24. While the first term of those equations represents the operating costs, the remaining two terms constitute the capital expenditure of the mitigation technologies. For capital expenditure, the second term relates to the fixed cost associated with the development of a new plant e.g., land and machinery. Meanwhile, the third term is the fixed cost associated with the plant's capacity. A larger plant capacity would have a higher fixed cost and vice versa.

$$\sum_i \left( (FNS_{i,k} OF_{i,k}) + (AFF . A_{i,k} FC1_{i,k}) + (AFF . FNS_{i,k} FC2_{i,k}) \right) = CTF_k \quad \forall k \quad \text{Equation 24}$$

where AFF is the annualized cost factor, $OF_{i,k}$, $FC1_{i,k}$ and $FC2_{i,k}$ are the operational costs, fixed capital costs and capacity-dependent capital costs of power plant $i$ in period $k$ respectively.

Following this, the total costs associated with the deployment of renewable energy source $r$ ($CTC_k$), EP-NETs technology $p$ ($CTEP_k$) and EC-NETs technology $q$ ($CTEQ_k$) in period $k$ are determined from Equation 25, Equation 26 and Equation 27 respectively.

$$\sum_r \left( (FC_{k,r} OC_{k,r}) + (AFF . C_{k,r} CC1_{k,r}) + (AFF . FC_{k,r} CC2_{k,r}) \right) = CTC_k \quad \forall k \quad \text{Equation 25}$$

$$\sum_p \left( (FEP_{k,p} OEP_{k,p}) + (AFF . D_{k,p} EPC1_{k,p}) + (AFF . FEP_{k,p} EPC2_{k,p}) \right) = CTEP_k \quad \forall k \quad \text{Equation 26}$$

$$\sum_q \left( (FEC_{k,q} OEC_{k,q}) + (AFF . E_{k,q} ECC1_{k,q}) + (AFF . FEC_{k,q} ECC2_{k,q}) \right) = CTEQ_k \quad \forall k \quad \text{Equation 27}$$

where $OC_{k,r}$, $OEP_{k,p}$ and $OEC_{k,q}$ are the operational costs of renewable energy source $r$, EP-NETs technology $p$ and EC-NETs technology $q$ in period $k$ respectively. Meanwhile, $CC1_{k,r}$, $EPC1_{k,p}$ and $ECC1_{k,q}$ are the fixed capital costs of renewable energy source $r$, EP-NETs technology $p$ and EC-NETs technology $q$ in period $k$ respectively. Also, $CC2_{k,r}$, $EPC2_{k,p}$ and $ECC2_{k,q}$ are the capacity-dependent capital costs of renewable energy source $r$, EP-NETs technology $p$ and EC-NETs technology $q$ in period $k$ respectively.

The total cost of the energy planning period $k$ ($TC_k$) is calculated from Equation 28. Note that this calculation procedure considers all power plants and mitigation technologies, including those calculated from Equation 25, Equation 26 and Equation 27.

$$CTF_k + \sum_i \sum_n (FNR_{i,k,n} \, CTR_{k,n} + AFF.CFR_{k,n} B_{i,k,n})$$
$$+ \sum_i \sum_s (FAS_{i,k,s} \, CTAS_{k,s} + AFF.CFAS_{k,s} G_{i,k,s})$$
$$+ \sum_i \sum_g (FAG_{i,k,g} \, CTAG_{k,g} + AFF.CFAG_{k,g} H_{i,k,g}) + CTC_k$$
$$+ CTEP_k + CTEC_k = TC_k \qquad \forall k \qquad \text{Equation 28}$$

where $CTR_{k,n}$ is the power generation cost with the deployment of CCS technology $n$ in period $k$ and $CFR_{k,n}$, $CFAS_{k,s}$ and $CFAG_{k,g}$ are the fixed costs associated with the deployment of CCS technology $n$, alternative solid-based fuel $s$ and gas-based fuels $g$ in period $k$ respectively. Meanwhile, $CTAS_{k,s}$ and $CTAG_{k,g}$ are the costs of alternative solid-based fuel $s$ and gas-based fuels $g$ in period $k$ respectively.

Subsequently, Equation 29 and Equation 30 present the constraints related to the total $CO_2$ emissions and total energy planning cost respectively.

$$TE_k \leq L_k \qquad \forall k \qquad \text{Equation 29}$$

$$TC_k \leq BD_k \qquad \forall k \qquad \text{Equation 30}$$

where $BD_k$ is the budget allocation of energy planning period $k$.

The objective function of this work is set to minimise either the total energy planning cost (Equation 31) or total $CO_2$ emissions (Equation 32). If the former is selected as the objective function, the constraints from Equation 1 to Equation 29 ensure that the $CO_2$ emissions limit of a geographical region in period $k$ is satisfied. Meanwhile, for the latter objective function, the constraints from Equation 1 to Equation 28 and Equation 30 limit the deployment of mitigation technologies subject to the budget availability of energy planning period $k$. Therefore, the $CO_2$ emissions limit may or may not be satisfied.

$$\min TC_k \qquad \text{Equation 31}$$

$$\min TE_k \qquad \text{Equation 32}$$

The presence of both continuous and integer variables results in the formulation being a MILP model. The mathematical formulation in this work is set up in Python, by using the open-source modelling language Pyomo (Bynum et al., 2017). The use of Pyomo allows users to freely use and modify the developed software. Additionally, this software may be utilised by all types of industries as it is in a publicly available domain. Meanwhile, an easy-to-use input spreadsheet was developed in Microsoft Excel for the inclusion of the necessary energy planning data. The optimisation problem is solved to global optimality by using the CPLEX solver from GAMS (GAMS, 2022), however, can also easily be solved using open-source solvers such as CBC and Octeract (Octeract Optimisation Intelligence, 2022). In other words, one may not need access to GAMS for the use of the DECO2 software. Note that both CBC and Octeract are available in the public domain. The mathematical formulation of this work titled **'Base_Model_Python.py'** is available on the *DECO2* GitHub page; accessible via https://github.com/mchlshort/DECO2. The next section presents the software infrastructure of the process integration-based software for optimal decarbonisation.

4. **Software Infrastructure**

The superstructural model formulation in this work is set up in Python, using the Pyomo algebraic modelling package, with an integrated user interface in Microsoft Excel; demonstrated in Figure 2.

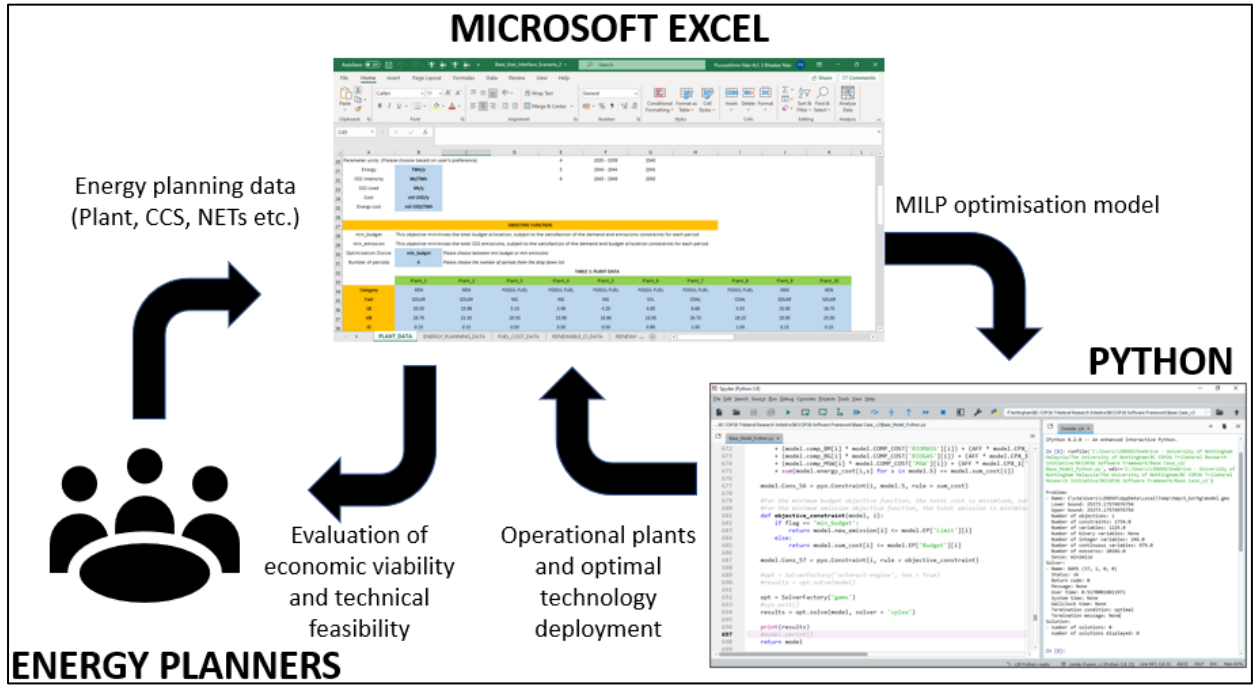

*Figure 2: Software infrastructure*

The energy planning data that is required during optimisation is imported from the user-interface file titled **'Base_User_Interface.xlsx'** which had been set up in Microsoft Excel. Besides creating a user-friendly tool, the inclusion of the energy planning data in a Microsoft Excel file allows energy planners without programming knowledge to utilise the software framework to carry out optimal decarbonisation. Figure 3 presents the snapshot of the energy planning data import in Python.

*Figure 3: Snapshot of energy planning data import in Python*

Note that Figure 3 presents the Spyder environment where the superstructural model of the software framework is formulated. Based on Figure 3, line 41 presents the name of the user-interface file which consists of the energy planning data. Note that a user needs only alter the entry on line 41 if the user-interface file is saved under a different name. Meanwhile, lines 42-56 represent the codes for the energy planning data import from the user-interface file. Once again, note that a user needs only change the entry of the *'sheet_name'* if the tab in Microsoft Excel is renamed. Figure 4 presents a snapshot of the Microsoft Excel-based user-interface file.

*Figure 4: Snapshot of the Microsoft Excel-based user-interface file*

Based on Figure 4, note that there are fifteen sets of data that must be included in the user-interface file. Each tab in the user-interface file consists of data that must be included by energy planners before optimising the superstructural model in Python. Table 2 presents the energy planning information related to each tab in the user-interface file.

*Table 2: Energy planning information related to each tab in the user-interface file*

| Microsoft Excel Tab | Energy Planning Information |
|---|---|
| PLANT_DATA | Type of fuels, lower and upper bounds of power generation, $CO_2$ intensities, commissioning, and decommissioning timeline of existing and upcoming power plants |
| ENERGY_PLANNING_DATA | Power demand, $CO_2$ emissions limit and budget availability in each energy planning period |
| FUEL_COST_DATA | Costs of fuels utilised in power plants in each energy planning period |
| RENEWABLE_CI_DATA | $CO_2$ intensities of available renewable energies in each energy planning period |
| RENEWABLE_COST_DATA | Costs of available renewable energies in each energy planning period |
| CAPEX_DATA_1 | Fixed capital costs of mitigation technologies in each energy planning period |
| CAPEX_DATA_2 | Capacity-dependent capital costs of mitigation technologies in each energy planning period |
| ALT_SOLID_CI | $CO_2$ intensities of alternative solid-based fuels in each energy planning period |
| ALT_SOLID_COST | Costs of alternative solid-based fuels in each energy planning period |
| ALT_GAS_CI | $CO_2$ intensities of alternative gas-based fuels in each energy planning period |
| ALT_GAS_COST | Costs of alternative gas-based fuels in each energy planning period |
| CCS_DATA | Removal ratios, parasitic power loss, power generation costs and fixed costs of CCS technologies in each energy planning period |
| NET_CI_DATA | $CO_2$ intensities of available NETs in each energy planning period |
| NET_COST_DATA | Costs of available NETs in each energy planning period |
| TECH_IMPLEMENTATION_TIME | The availabilities of mitigation technologies in each energy planning period |

As indicated in Table 2, the entries for all tabs (except *'TECH_IMPLEMENTATION_TIME'*) in the user-interface file consist of numerical values. However, the information in the *'TECH_IMPLEMENTATION_TIME'* tab is non-numerical. Instead, an energy planner needs to input the availabilities of the mitigation technologies in each period in terms of 'YES' (present) and 'NO' (absent), as shown in Figure 5.

|   | A | B | C | D | E | F | G | H | I | J |
|---|---|---|---|---|---|---|---|---|---|---|
| 18 |   |   |   |   |   |   |   |   |   |   |
| 19 |   |   |   |   |   |   |   |   | TABLE 14: TECHNOLOGY IMPLEMENTATION TIME | |
| 20 | Period | SOLID_1 | SOLID_2 | GAS_1 | GAS_2 | CCS_1 | CCS_2 | SOLAR | HYDRO | BIOMASS |
| 21 | 1 | NO | NO | NO | NO | NO | NO | YES | YES | NO |
| 22 | 2 | NO | NO | NO | NO | NO | NO | YES | YES | NO |
| 23 | 3 | YES | NO | YES | NO | NO | NO | YES | YES | YES |
| 24 | 4 | YES | YES | YES | YES | YES | YES | YES | YES | YES |
| 25 | 5 | YES | YES | YES | YES | YES | YES | YES | YES | YES |
| 26 | 6 | YES | YES | YES | YES | YES | YES | YES | YES | YES |

*Figure 5: Information for the 'TECH_IMPLEMENTATION_TIME' tab in the user-interface file*

For example, for the case in Figure 5, only solar and hydropower mitigation technologies are available in Period 1 whilst all mitigation technologies are available in Period 6. Therefore, only solar and hydropower mitigation technologies may be used to mitigate the $CO_2$ emissions in Period 1. The absence of other mitigation technologies may be due to factors such as the lack of maturity with the associated technologies, policy, or implementation time constraints. Upon inputting all energy planning data in the user-interface file, an energy planner may choose the objective function from the *'PLANT_DATA'* tab, as shown in Figure 6.

|    | A | B | C | D | E | F | G | H | I |
|----|---|---|---|---|---|---|---|---|---|
| 16 |   |   | User to define heading for each plant | | Period | Years | Data year |   |   |
| 17 |   |   | User to define parameters for each plant/period | | 1 | 2020 - 2024 | 2025 |   |   |
| 18 |   |   | User should not alter these headings | | 2 | 2025 - 2029 | 2030 |   |   |
| 19 |   |   |   |   | 3 | 2030 - 2034 | 2035 |   |   |
| 20 | Parameter units  (Please choose based on user's preference) | | | | 4 | 2035 - 2039 | 2040 |   |   |
| 21 | Energy | TWh/y |   |   | 5 | 2040 - 2044 | 2045 |   |   |
| 22 | CO2 Intensity | Mt/TWh |   |   | 6 | 2045 - 2049 | 2050 |   |   |
| 23 | CO2 Load | Mt/y |   |   |   |   |   |   |   |
| 24 | Cost | mil USD/y |   |   |   |   |   |   |   |
| 25 | Energy cost | mil USD/TWh |   |   |   |   |   |   |   |
| 26 |   |   |   |   |   |   |   |   |   |
| 27 | OBJECTIVE FUNCTION | | | | | | | | |
| 28 | min_budget | This objective minimises the total budget allocation, subject to the satisfaction of the demand and emissions constraints for each period | | | | | | | |
| 29 | min_emission | This objective minimises the total CO2 emissions, subject to the satisfaction of the demand and budget allocation constraints for each period | | | | | | | |
| 30 | Optimisation Choice | min_budget | Please choose between min budget or min emissions | | | | | | |
| 31 | Number of periods | 6 | Please choose the number of periods from the drop down list | | | | | | |

*Figure 6: Snapshot of the 'PLANT_DATA' tab*

Based on Figure 6, cell 'B30' presents the choices of objective functions available in this work. The available objective functions are *'min_budget'* (Equation 31) and *'min_emission'* (Equation 32). An energy planner may choose one of the objective functions depending on one's preference. Also, the energy planner may specify the number of energy planning periods based on the dropdown list in cell 'B31'. An energy planner may choose between 1 and 50 periods. Depending on the number of energy planning periods selected, an energy planner should input the energy planning data for the specified number of periods. In Figure 5, the energy planning data for the **'TECH_IMPLEMENTATION_TIME'** tab is specified for six periods.

Next, an energy planner may optimise the superstructural model in the Python file titled *'Base_Model_Python'*. Lines 692 and 694 of Figure 7 present the solver statement of the superstructural model which makes use of the CPLEX solver from GAMS (GAMS, 2022). Note that a user may choose to alter the solver name on line 694 to any other suitable MILP solver if one does not have access to the CPLEX solver.

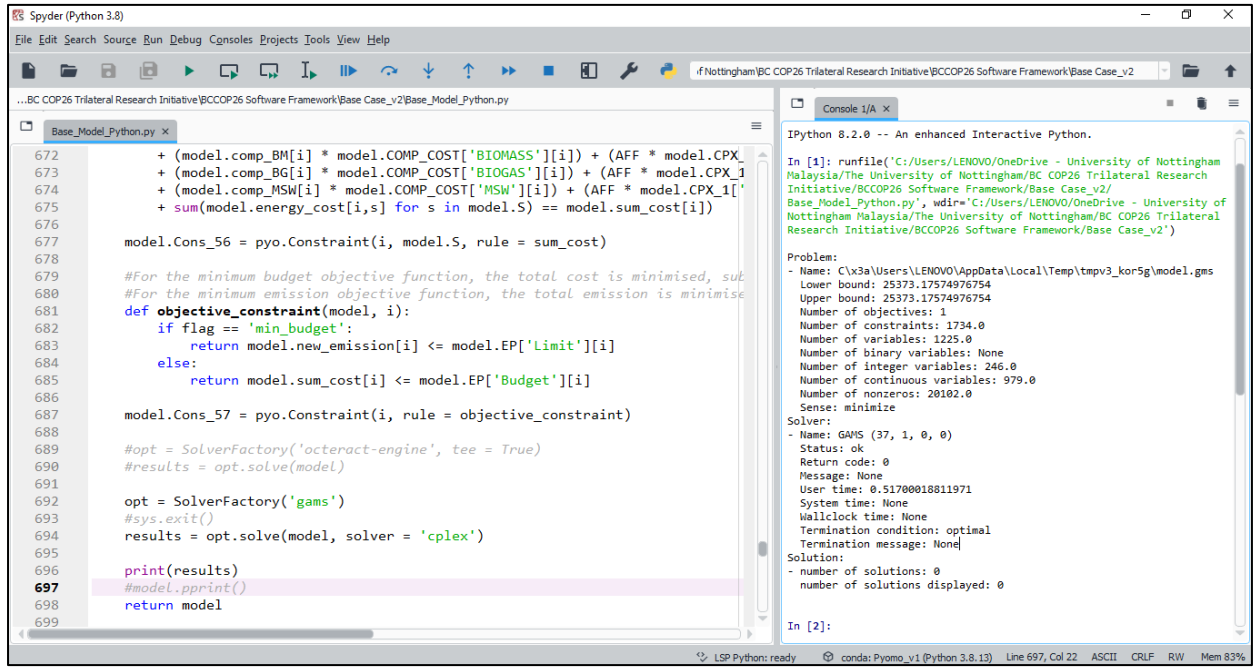

*Figure 7: Optimising the superstructural model in Python*

Note that the right side of Figure 7 represents the console of the Spyder environment. Upon optimising the superstructural model, the console displays the solver status and parameters associated with an optimisation problem e.g., number of objectives, variables, user time etc. If the termination condition is mentioned as 'optimal', it indicates that the superstructural model is solved to global optimality. An energy planner may now re-open the Microsoft Excel file for results viewing. Figure 8 presents the snapshot of the results in the user-interface file.

Since six periods were chosen previously in Figure 6, an equivalent number of tabs were created in the user-interface file, as shown in Figure 8. The results in Figure 8 consist of all power plants specified in the **'PLANT_DATA'** tab, as well as the available mitigation technologies. Table 3 describes the definition of the column heading in Figure 8.

At this stage, an energy planner may analyse and evaluate the results to determine the practicality and ease of deployment of mitigation technologies. If any results are deemed unsuitable or overly optimistic, an energy planner may alter the energy planning data before re-running the optimisation software. Note that the energy planning may only need to alter the energy planning data since all variables and constraints had been specified in Python with no inputs required from an energy planner.

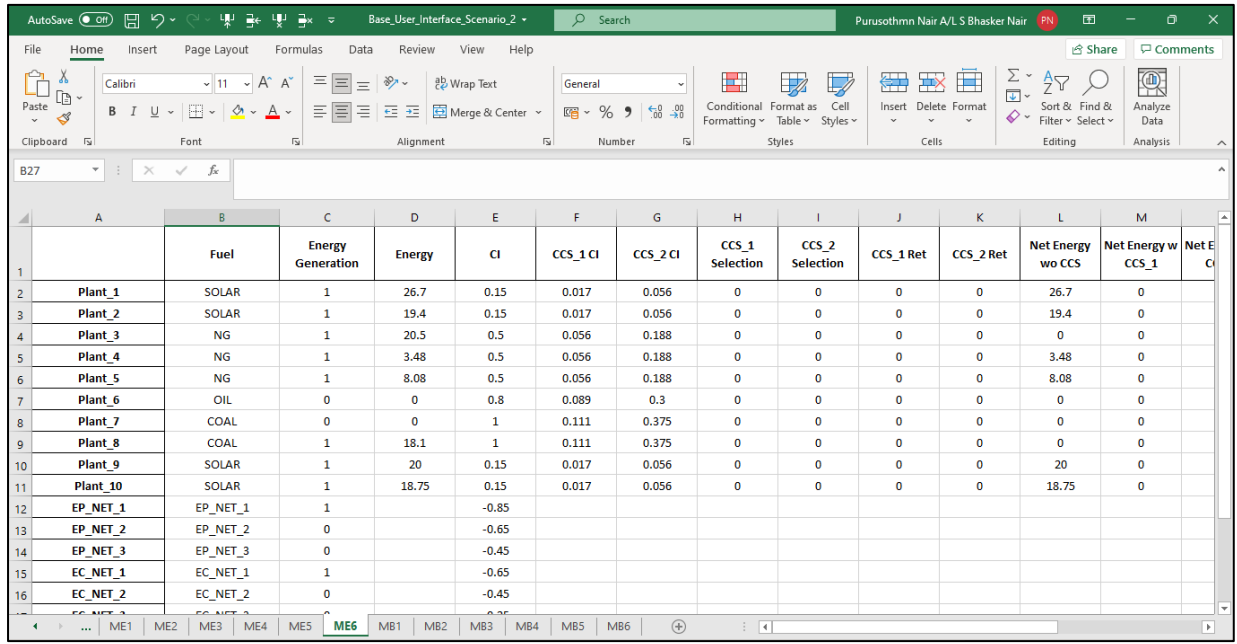

*Figure 8: Results snapshot of the optimised superstructural model*

Table 3: Description of the column heading under the results tab

| Column heading | Description |
|---|---|
| Fuel | Type of fuel used in the power plant |
| Gross Energy | Gross energy generated by each power plant |
| CCS_1 Ret | Energy from each power plant subjected to the deployment of CCS technology 1 |
| CCS_2 Ret | Energy from each power plant subjected to the deployment of CCS technology 2 |
| SOLID_1 | Energy generation by alternative solid-based fuel type 1 in each power plant |
| SOLID_2 | Energy generation by alternative solid-based fuel type 2 in each power plant |
| GAS_1 | Energy generation by alternative gas-based fuel type 1 in each power plant |
| GAS_2 | Energy generation by alternative gas-based fuel type 2 in each power plant |
| Net Energy | The net energy of each power plant and mitigation technology (NETs & renewable energy) |
| $CO_2$ Load | The total $CO_2$ load of each power plant and mitigation technology (NETs & renewable energy) |
| Cost | The total cost of the energy planning system |

The next section presents the case study that is used to demonstrate the optimal decarbonisation software framework.

## 5. Case Study

A hypothetical case study is used to demonstrate the application of the process-integrated based software framework for optimal decarbonisation. The superstructural model in this work is demonstrated with six periods, each spanning a time interval of five years. Table 4 presents the energy planning data that is specified in the user interface file titled **'Base_User_Interface.xlsx'**. Note that the data in Table 4 are assumed and do not represent any real-life scenario. These energy planning data though not representative of an industrial scenario would demonstrate the applicability of the software for optimal decarbonisation. The data in Table 4 are arranged in the order of the Microsoft Excel tabs mentioned in Table 2. Note that the data relevant to the **'TECH_IMPLEMENTATION_TIME'** tab is excluded for now.

*Table 4 (a): Case study energy planning data: 'PLANT_DATA'*

| Plant | Category | Fuel | Lower Bound (TWh y$^{-1}$) | Upper Bound (TWh y$^{-1}$) | CO$_2$ Intensity (Mt TWh$^{-1}$) | $CM_i$ | $DCM_i$ |
|---|---|---|---|---|---|---|---|
| **Plant 1** | Renewable | Solar | 20.03 | 26.70 | 0.15 | 1 | 7 |
| **Plant 2** | | | 15.98 | 21.30 | 0.15 | 1 | 7 |
| **Plant 3** | Fossil Fuel | Natural Gas | 5.13 | 20.50 | 0.50 | 1 | 7 |
| **Plant 4** | | | 3.48 | 13.90 | 0.50 | 1 | 7 |
| **Plant 5** | | | 4.20 | 16.80 | 0.50 | 1 | 7 |
| **Plant 6** | | Oil | 4.00 | 16.00 | 0.80 | 1 | 3 |
| **Plant 7** | | Coal | 6.68 | 26.70 | 1.00 | 1 | 5 |
| **Plant 8** | | | 4.53 | 18.10 | 1.00 | 1 | 7 |
| **Plant 9** | Renewable | Solar | 15.00 | 20.00 | 0.15 | 2 | 7 |
| **Plant 10** | | | 18.75 | 25.00 | 0.15 | 4 | 7 |

As shown in Table 4a, there are 10 power plants available for power generation. While the first eight plants are existing power plants (commissioned from Period 1 onwards), Plants 9 and 10 are upcoming power plants to be commissioned from Periods 2 and 4 respectively. In other words, Plants 9 and 10 would not be available for power generation before Periods 2 and 4, respectively. On the other hand, Plants 6 and 7 utilising oil and coal respectively would be decommissioned in Periods 3 and 5 respectively. Meanwhile, the lower bound of power generation by plants utilising renewable energy sources (Plants 1, 2, 9 and 10) is set to 75% of their upper bounds. In other words, the power generation from operational renewable-based power plants should at least be 75% of its maximum design capacity since these plants cannot be ramped up easily to meet a sudden demand surge (Impram et al., 2020). By contrast, power generation from plants utilising fossil-based sources may be ramped up quickly. Therefore, the lower bound for these plants (Plants 3 till 8) is set to 25% of their maximum generation capacity.

*Table 5 (b): Case study energy planning data: 'ENERGY_PLANNING_DATA'*

| Energy planning parameters | 1 | 2 | 3 | 4 | 5 | 6 |
|---|---|---|---|---|---|---|
| Demand (TWh $y^{-1}$) | 60 | 75 | 90 | 105 | 120 | 135 |
| $CO_2$ Emissions Limit (Mt) | 20 | 18 | 15 | 11 | 6 | 0 |
| Budget (mil USD $y^{-1}$) | 3,000 | 3,500 | 4,000 | 4,500 | 5,000 | 5,500 |

*Table 6 (c): Case study energy planning data: 'FUEL_COST_DATA'*

| Fuel cost (mil USD $TWh^{-1}$) | 1 | 2 | 3 | 4 | 5 | 6 |
|---|---|---|---|---|---|---|
| Natural Gas | 25 | | | | | |
| Oil | 49 | | | | | |
| Coal | 12 | | | | | |
| Solar | 40 | 35 | 25 | 13 | 8 | 3 |

In Table 4b, the energy planning is conducted based on incremental energy demand and stricter $CO_2$ emissions limits between successive periods. This scenario is often observed in developing countries with increasing populations which leads to higher demand requirements. At the same time, lower $CO_2$ emissions limits are required for countries to meet their climate change targets. Note that the $CO_2$ emissions limit is observed to undergo a drastic decline at later periods. It is projected that the greater availability of mitigation technologies at later periods would make it relatively easier to drive $CO_2$ emissions reduction. Note that this work targets to achieve net-zero emissions by the final period, consistent with the pledges made at the COP26 (United Nations Climate Change, 2021b). Also, economic growth contributes to a greater budget available for energy planning across periods. Since the power plants in Table 4a make use of solar, natural gas, oil and coal, the associated fuel costs are presented in Table 4c. As fossil-based sources are technologically matured, their costs are projected to remain constant for all periods (Ali et al., 2021; Riusxander and Sumirat, 2013). By contrast, recent technological advancement has resulted in a significant decline in the cost of solar energy (Armstrong, 2021; ASEAN Centre for Energy, 2016; Green, 2019). This information is captured in this work with the declining cost of solar energy in Table 4c. Note that the cost decline in earlier periods is gradual before increasing drastically towards later periods.

Aside from the existing plants, this work also considers the potential deployment of renewable energies as separate plants for the mitigation of $CO_2$ emissions. The five renewable energies that are considered in this work are solar, hydropower, biomass, biogas, and municipal solid waste (MSW). Each renewable energy differs in terms of $CO_2$ intensities (Table 4d) and costs (Table 4e). Like solar energy, the $CO_2$ intensities and costs of renewable energies are expected to decrease across periods (Green, 2019; IRENA, 2021). In the final period, solar energy would be the cheapest (Armstrong, 2021; Green, 2019) and has the lowest $CO_2$ intensity (Nair et al., 2021; Sahu et al., 2014). The superstructural model in this work would determine the optimal deployment of renewable energy sources (if necessary) to meet the demand and $CO_2$ emissions limits.

*Table 7 (d): Case study energy planning data: 'COMPENSATORY_CI_DATA'*

| $CO_2$ intensity of renewable energy (Mt TWh$^{-1}$) | 1 | 2 | 3 | 4 | 5 | 6 |
|---|---|---|---|---|---|---|
| Solar | 0.10 | 0.09 | 0.08 | 0.07 | 0.06 | 0.05 |
| Hydropower | 0.15 | 0.14 | 0.13 | 0.12 | 0.11 | 0.10 |
| Biomass | 0.30 | 0.28 | 0.26 | 0.24 | 0.22 | 0.20 |
| Biogas | 0.25 | 0.23 | 0.21 | 0.19 | 0.17 | 0.15 |
| Municipal Solid Waste | 0.30 | 0.29 | 0.28 | 0.27 | 0.26 | 0.25 |

*Table 8 (e): Case study energy planning data: 'COMPENSATORY_COST_DATA'*

| Cost of renewable energy (mil USD TWh$^{-1}$) | 1 | 2 | 3 | 4 | 5 | 6 |
|---|---|---|---|---|---|---|
| Solar | 40 | 35 | 25 | 13 | 8 | 3 |
| Hydropower | 30 | 29 | 28 | 27 | 26 | 25 |
| Biomass | 20 | 18 | 16 | 14 | 12 | 10 |
| Biogas | 25 | 23 | 21 | 19 | 17 | 15 |
| Municipal Solid Waste | 20 | 19 | 18 | 17 | 16 | 15 |

Next, Table 4f and Table 4g present the capital costs associated with all plants (fossil fuel, renewable and NETs). Note that this work assumes the capital expenditure for NETs plants to be higher than for fossil-based plants. Given that NETs are still in an early development phase with the lack of technological maturity, it is assumed that greater initial investment is required for NETs plants. However, the capital costs of all plants except fossil-based plants are expected to decrease across periods, with solar witnessing the largest decline. The capital costs for fossil-based plants are projected to remain constant and their values are derived from the work of Ali et al. (2021) and Riusxander and Sumirat (2013).

In this work, alternative solid (biomass) and gas-based fuels (biogas) are meant to replace coal and natural gas respectively. For each fuel replacement, there are two types available for use. For example, the two choices of biomass could be empty fruit bunch (EFB) and palm kernel shell (PKS) (Kaniapan et al., 2021). Meanwhile, examples of biogas may be palm oil mill effluent (POME) (Kaniapan et al., 2021) and animal manure (Ramos-Suárez et al., 2019). In this work, it is assumed that the cheaper alternative fuel would have higher $CO_2$ intensity and vice versa. The reasoning behind this assumption is that fuels with lower $CO_2$ intensity would often be subjected to processes with high operating costs. The $CO_2$ intensities and cost of the alternative fuels are presented in Table 4h - Table 4k. Once again, the improved energy efficiencies are projected to decrease the $CO_2$ intensities and costs of alternative fuels across periods.

*Table 9 (f): Case study energy planning data: 'CAPEX_DATA_1'*

| Fixed capital costs (mil USD TWh$^{-1}$) | 1 | 2 | 3 | 4 | 5 | 6 |
|---|---|---|---|---|---|---|
| Natural Gas | | | | | | |
| Oil | | | 400 | | | |
| Coal | | | | | | |
| Solar | 400 | 350 | 300 | 250 | 200 | 150 |
| Hydropower | 400 | 380 | 360 | 340 | 320 | 300 |
| Biogas | | | | | | |
| Biomass | 400 | 390 | 380 | 370 | 360 | 350 |
| Municipal Solid Waste | | | | | | |
| EP-NETs technology 1 | | | | | | |
| EP-NETs technology 2 | 600 | 550 | 500 | 450 | 400 | 350 |
| EP-NETs technology 3 | | | | | | |
| EC-NETs technology 1 | | | | | | |
| EC-NETs technology 2 | 800 | 750 | 700 | 650 | 600 | 550 |
| EC-NETs technology 3 | | | | | | |

*Table 10 (g): Case study energy planning data: 'CAPEX_DATA_2'*

| Fixed capital costs (mil USD TWh$^{-1}$) | 1 | 2 | 3 | 4 | 5 | 6 |
|---|---|---|---|---|---|---|
| Natural Gas | | | | | | |
| Oil | | | 100 | | | |
| Coal | | | | | | |
| Solar | 100 | 85 | 70 | 55 | 40 | 25 |
| Hydropower | 100 | 90 | 80 | 70 | 60 | 50 |
| Biogas | | | | | | |
| Biomass | 100 | 95 | 90 | 85 | 80 | 75 |
| Municipal Solid Waste | | | | | | |
| EP-NETs technology 1 | | | | | | |
| EP-NETs technology 2 | 150 | 140 | 130 | 120 | 110 | 100 |
| EP-NETs technology 3 | | | | | | |
| EC-NETs technology 1 | | | | | | |
| EC-NETs technology 2 | 200 | 190 | 180 | 170 | 160 | 150 |
| EC-NETs technology 3 | | | | | | |

*Table 11 (h): Case study energy planning data: 'ALT_SOLID_CI'*

| CO$_2$ intensity of alternative solid-based fuel (Mt TWh$^{-1}$) | 1 | 2 | 3 | 4 | 5 | 6 |
|---|---|---|---|---|---|---|
| Technology 1 | 0.20 | 0.19 | 0.18 | 0.17 | 0.16 | 0.15 |
| Technology 2 | 0.40 | 0.38 | 0.36 | 0.34 | 0.32 | 0.30 |

*Table 12 (i): Case study energy planning data: 'ALT_SOLID_COST'*

| Cost of alternative solid-based fuel (Mt TWh$^{-1}$) | 1 | 2 | 3 | 4 | 5 | 6 |
|---|---|---|---|---|---|---|
| Technology 1 | 20 | 19 | 18 | 17 | 16 | 15 |
| Technology 2 | 15 | 14 | 13 | 12 | 11 | 10 |

*Table 13 (j): Case study energy planning data: 'ALT_GAS_CI'*

| CO$_2$ intensity of alternative gas-based fuel (Mt TWh$^{-1}$) | 1 | 2 | 3 | 4 | 5 | 6 |
|---|---|---|---|---|---|---|
| Technology 1 | 0.15 | 0.14 | 0.13 | 0.12 | 0.11 | 0.10 |
| Technology 2 | 0.25 | 0.23 | 0.21 | 0.19 | 0.17 | 0.15 |

*Table 14 (k): Case study energy planning data: 'ALT_GAS_COST'*

| Cost of alternative gas-based fuel (Mt TWh$^{-1}$) | 1 | 2 | 3 | 4 | 5 | 6 |
|---|---|---|---|---|---|---|
| Technology 1 | 35 | 34 | 33 | 32 | 31 | 30 |
| Technology 2 | 30 | 29 | 28 | 27 | 26 | 25 |

Following this, the CCS data is presented in Table 4l. Note that there are two CCS technologies available for deployment. Examples of CCS technologies are pre-combustion, post-combustion and oxyfuel capture (Gibbins and Chalmers, 2008). In this work, these two CCS technologies may represent any of the available technologies. CCS technology 1 (e.g., pre-combustion capture) has a higher removal ratio and lower parasitic power loss in comparison to CCS technology 2 (e.g., post-combustion capture) (Kheirinik et al., 2021; Tock and Maréchal, 2013). Therefore, the latter has a lower cost in comparison to CCS technology 1. Due to the projected improvement in the technological maturity of CCS, the removal ratios of CCS systems are expected to increase. Meanwhile, the remaining CCS parameters (parasitic power loss and costs) are projected to decline across periods.

Finally, the CO$_2$ intensities and costs of NETs are presented in Table 4m and Table 4n respectively. Note that both EP-NETs and EC-NETs are considered in this work, where each NETs type is made up of three technologies. Some examples of EP-NETs are biochar and BECCS, while EC-NETs are made up of DAC and enhanced weathering (EW). Note that NETs with the lowest CO$_2$ intensity (highest CDR capability) is the most expensive technology and vice versa. Once again, like CCS technologies and renewable energy, all

NETs are projected to mature across periods resulting in declining $CO_2$ intensities and costs. The $CO_2$ intensities and costs of NETs are derived from the work of Keith et al. (2018) and Strefler et al. (2018).

*Table 15 (l): Case study energy planning data: ''CCS_DATA'*

| CCS data | 1 | 2 | 3 | 4 | 5 | 6 |
|---|---|---|---|---|---|---|
| Removal ratio of CCS technology 1 | 0.85 | 0.86 | 0.87 | 0.88 | 0.89 | 0.90 |
| Parasitic power loss of CCS technology 1 | 0.15 | 0.14 | 0.13 | 0.12 | 0.11 | 0.10 |
| Power generation cost of CCS technology 1 (mil USD TWh$^{-1}$) | 34 | 33 | 32 | 31 | 30 | 29 |
| Fixed cost of CCS technology 1 (mil USD TWh$^{-1}$) | 600 | 590 | 580 | 570 | 560 | 550 |
| Removal ratio of CCS technology 1 | 0.65 | 0.66 | 0.67 | 0.68 | 0.69 | 0.70 |
| Parasitic power loss of CCS technology 1 | 0.25 | 0.24 | 0.23 | 0.22 | 0.21 | 0.20 |
| Power generation cost of CCS technology 2 (mil USD TWh$^{-1}$) | 29 | 28 | 27 | 26 | 25 | 24 |
| Fixed cost of CCS technology 2 (mil USD TWh$^{-1}$) | 550 | 540 | 530 | 520 | 510 | 500 |

*Table 16 (m): Case study energy planning data: 'NET_CI_DATA'*

| $CO_2$ intensity of NETs (Mt TWh$^{-1}$) | 1 | 2 | 3 | 4 | 5 | 6 |
|---|---|---|---|---|---|---|
| EP-NETs technology 1 | -0.80 | -0.81 | -0.82 | -0.83 | -0.84 | -0.85 |
| EP-NETs technology 2 | -0.60 | -0.61 | -0.62 | -0.63 | -0.64 | -0.65 |
| EP-NETs technology 3 | -0.40 | -0.41 | -0.42 | -0.43 | -0.44 | -0.45 |
| EC-NETs technology 1 | -0.60 | -0.61 | -0.62 | -0.63 | -0.64 | -0.65 |
| EC-NETs technology 2 | -0.40 | -0.41 | -0.42 | -0.43 | -0.44 | -0.45 |
| EC-NETs technology 3 | -0.20 | -0.21 | -0.22 | -0.23 | -0.24 | -0.25 |

*Table 17 (n): Case study energy planning data: 'NET_COST_DATA'*

| Cost of NETs (mil USD TWh$^{-1}$) | 1 | 2 | 3 | 4 | 5 | 6 |
|---|---|---|---|---|---|---|
| EP-NETs technology 1 | 43 | 41 | 39 | 37 | 35 | 33 |
| EP-NETs technology 2 | 40 | 38 | 36 | 34 | 32 | 30 |
| EP-NETs technology 3 | 37 | 35 | 33 | 31 | 29 | 27 |
| EC-NETs technology 1 | 49 | 47 | 45 | 43 | 41 | 39 |
| EC-NETs technology 2 | 37 | 35 | 33 | 31 | 29 | 27 |
| EC-NETs technology 3 | 24 | 22 | 20 | 18 | 16 | 14 |

Two scenarios are evaluated in this work using the hypothetical case study. The first scenario is considered the less ambitious approach toward mitigating $CO_2$ emissions. By contrast, the second scenario is more aggressive and ambitious in addressing climate change issues. The next section presents the details of Scenario 1.

**5.1 Scenario 1**

In Scenario 1, only certain mitigation technologies are available for $CO_2$ emissions mitigation, given as in Table 18.

*Table 18: Availability of mitigation technologies in Scenario 1*

| Technology Availability | 1 | 2 | 3 | 4 | 5 | 6 |
|---|---|---|---|---|---|---|
| Solar | ✓ | ✓ | ✓ | ✓ | ✓ | ✓ |
| Hydropower | ✓ | ✓ | ✓ | ✓ | ✓ | ✓ |
| Biomass | ✓ | ✓ | ✓ | ✓ | ✓ | ✓ |
| Biogas | ✓ | ✓ | ✓ | ✓ | ✓ | ✓ |
| MSW | ✗ | ✗ | ✗ | ✓ | ✓ | ✓ |
| Alternative solid-based fuel technology 1 | ✗ | ✗ | ✗ | ✗ | ✗ | ✗ |
| Alternative solid-based fuel technology 2 | ✗ | ✗ | ✓ | ✓ | ✓ | ✓ |
| Alternative gas-based fuel technology 1 | ✗ | ✗ | ✗ | ✗ | ✗ | ✗ |
| Alternative gas-based fuel technology 2 | ✗ | ✗ | ✓ | ✓ | ✓ | ✓ |
| CCS technology 1 | ✗ | ✗ | ✗ | ✗ | ✗ | ✗ |
| CCS technology 2 | ✗ | ✗ | ✗ | ✓ | ✓ | ✓ |
| EP-NETs technology 1 | ✗ | ✗ | ✗ | ✗ | ✗ | ✗ |
| EP-NETs technology 2 | ✗ | ✗ | ✗ | ✗ | ✗ | ✗ |
| EP-NETs technology 3 | ✗ | ✗ | ✗ | ✗ | ✗ | ✗ |
| EC-NETs technology 1 | ✗ | ✗ | ✗ | ✗ | ✗ | ✗ |
| EC-NETs technology 2 | ✗ | ✗ | ✗ | ✗ | ✗ | ✗ |
| EC-NETs technology 3 | ✗ | ✗ | ✗ | ✗ | ✗ | ✗ |

Based on Table 18, all renewable energies (except MSW) are available for use in all energy planning periods. Note however that MSW is only available from Period 4. It is assumed that technology associated with MSW would take a longer time to mature (Perrot and Subiantoro, 2018). Besides, one alternative solid and gas-based fuel technology is unavailable in Scenario 1. These technologies are assumed to have lower $CO_2$ intensities (see Table 4h and Table 4j,), and are far more mature as compared to their counterparts (technologies 2). Therefore, only the less matured technologies of the alternative fuels are available for Scenario 1 and are only available from Period 3 onwards. The fuel substitution of coal and natural gas power plants would involve co-firing, thus requiring retrofit to be carried out on both types of power plants. Retrofitting power plants is capital-intensive and may not be available in earlier periods. Similar reasoning as to alternative fuels is applied for CCS. For the latter mitigation technology, only technology 2 is available from Period 4 onwards. Since CCS deployment is capital intensive, only one technology that has a lower removal ratio and higher parasitic power loss is available in Scenario 1. Finally, although NETs deployment is crucial for CDR, its commercial deployment is not likely to be seen anytime soon (Sandrine, 2021). Therefore, the deployment of NETs is absent in Scenario 1. The information in Table 18 is inputted in the **'TECH_IMPLEMENTATION_TIME'** tab in the user interface file titled **'Base_User_Interface.xlsx'**.

### 5.1.1 Case 1

Once all energy planning information is included in the user interface file, optimisation is carried out using a superstructural model coded in the Python file entitled **'Base_Model_Python'**. The objective function of Case 1 is set to minimise the total $CO_2$ emissions (Equation 32). In other words, the superstructural model is optimised subject to the budget availability in Table 4b. Note that **'min_emission'** is selected in cell 'B30' of the user-interface file (see Figure 6). Figure 9 presents the results of Case 1. The detailed results of Case 1 in Scenario 1 are presented in Table S1 in the Supplementary Information.

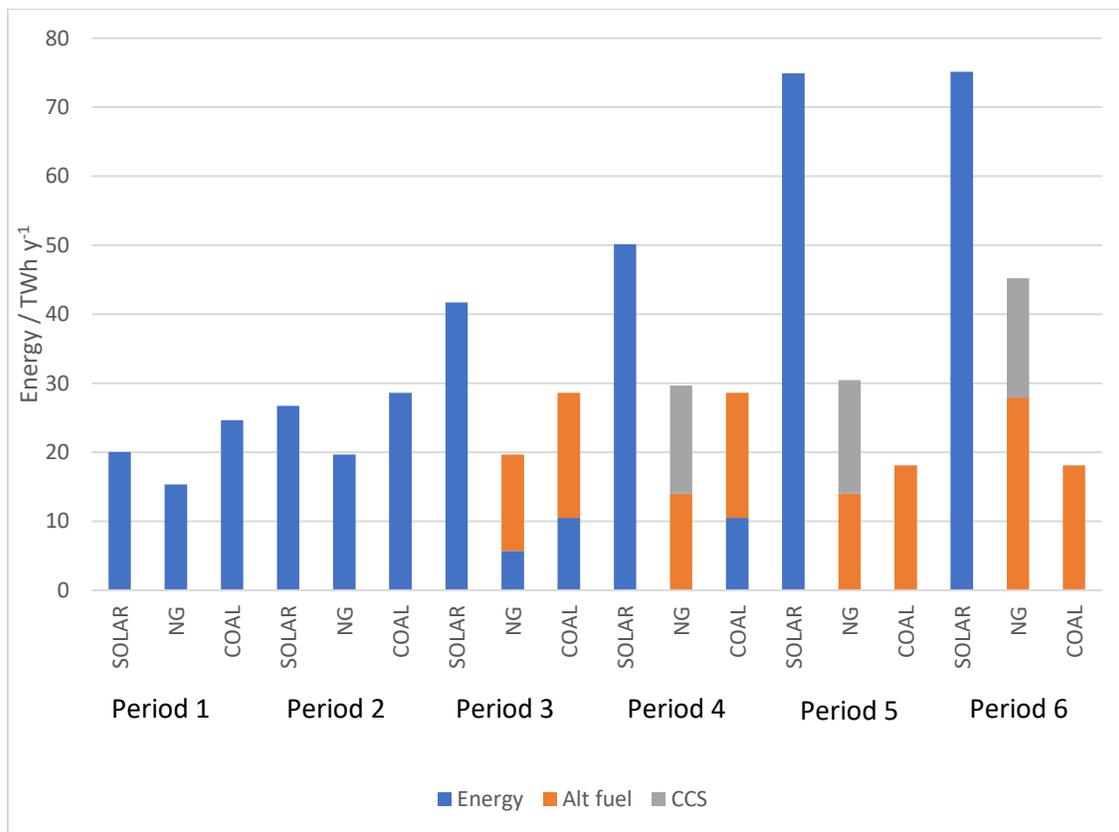

*Figure 9: Power plants configurations of Case 1 in Scenario 1*

Based on Figure 9, only power plants 1, 3, 7 and 8 were selected for power generation in Period 1. Among them, only one of them (power plant 1) is a solar plant (renewable energy). Although the power demand of 60 TWh y$^{-1}$ was satisfied, the total $CO_2$ emissions were minimised at 35 Mt y$^{-1}$, thus violating the $CO_2$ emissions limit of 20 Mt y$^{-1}$. Moving on to Period 2, similar configurations as to Period 1 were observed, with additional power generation from the existing operational power plants. Once again, the power demand was satisfied but the $CO_2$ emissions were violated by 24 Mt y$^{-1}$ (= 42 – 18 Mt y$^{-1}$). The commissioning of solar-based power plant 9 in Period 2 meant that it is available for power generation in Period 3, and thus selected to satisfy the demand of 90 TWh y$^{-1}$. In addition to that, both biomass and biogas were deployed to replace the fuels in power plants 3 (natural gas) and 8 (coal) respectively. The alternative fuels have lower $CO_2$ intensities, thus contributing to lower cumulative $CO_2$ emissions of 29 Mt y$^{-1}$. Despite the higher costs of alternative fuels, their deployment is relatively cheaper than commissioning new power plants. Nevertheless, the $CO_2$ emissions limit was violated by 14 Mt y$^{-1}$ (= 29 – 15 Mt y$^{-1}$).

In Period 4, power plant 5 fueled by natural gas was deployed to meet the incremental demand increase. However, its high $CO_2$ intensity increased the total $CO_2$ load, thus demanding additional mitigation measures. Therefore, both power plants 3 and 5 were retrofitted with CCS. The reduced power generation from those power plants is due to parasitic power losses during CCS deployment. Note that the CCS deployment in power plant 3 is in addition to its deployment of alternative gas-based fuel (biogas) technology 2. In other words, there were two mitigation technologies deployed for power plant 3. In addition to that, a new 3.45 TWh y$^{-1}$ solar-based power plant which has a $CO_2$ intensity of 0.07 Mt TWh$^{-1}$ was commissioned in Period 4. Note that the solar-based power plant 9 operated at its maximum capacity in Period 4. Despite additional mitigation measures and budget, the total $CO_2$ emissions in Period 4 were similar to Period 3. With the available budget, the total $CO_2$ could only be minimised to 29 Mt y$^{-1}$, thus violating the $CO_2$ emissions limit by 18 Mt y$^{-1}$ (= 29 – 11 Mt y$^{-1}$).

Moving on to Period 5, power plant 7 which generated power from Period 1 was decommissioned (see Table 4a). Therefore, power plant 10 powered by solar was deployed to meet the incremental demand. Note that solar power plants 1, 9 and 10 were operating at their maximum capacity. The deployment of mitigation technologies (CCS and alternative fuels) in Period 5 was similar to Period 4. A similar situation was observed in the final period, except that power plant 4 fueled by natural gas was now deployed to satisfy the demand of 135 TWh y$^{-1}$. Note that CCS was deployed in power plant 4 for the mitigation of its emissions. Despite the deployment of several mitigation technologies, they were insufficient to satisfy the $CO_2$ emissions limits in all periods in Scenario 1.

### 5.1.2 Case 2

In Case 2, the total energy planning cost (Equation 31) is minimised as the objective function. In other words, the $CO_2$ emissions limit for all periods must be satisfied. However, insufficient mitigation technologies (especially NETs) have resulted in an infeasible solution. In other words, CDR via NETs is necessary for the achievement of the net-zero target in the final energy planning period. Unless additional mitigation technologies are available, the $CO_2$ emissions limits in Scenario 1 would be constantly violated. Therefore, Scenario 2 is next investigated to identify the impact of additional mitigation technologies during energy planning.

### 5.2 Scenario 2

Scenario 2 is regarded to be more aggressive in comparison to Scenario 1, where all mitigation technologies are now available for deployment (see Table 19). Unlike Scenario 1, some technologies are available at earlier periods due to an assumption of rapid technology maturity.

*Table 19: Availability of mitigation technologies in Scenario 2*

| Technology Availability | 1 | 2 | 3 | 4 | 5 | 6 |
|---|---|---|---|---|---|---|
| Solar | ✓ | ✓ | ✓ | ✓ | ✓ | ✓ |
| Hydropower | ✓ | ✓ | ✓ | ✓ | ✓ | ✓ |
| Biomass | ✓ | ✓ | ✓ | ✓ | ✓ | ✓ |
| Biogas | ✓ | ✓ | ✓ | ✓ | ✓ | ✓ |
| MSW | ✗ | ✓ | ✓ | ✓ | ✓ | ✓ |
| Alternative solid-based fuel technology 1 | ✗ | ✗ | ✓ | ✓ | ✓ | ✓ |
| Alternative solid-based fuel technology 2 | ✗ | ✓ | ✓ | ✓ | ✓ | ✓ |
| Alternative gas-based fuel technology 1 | ✗ | ✗ | ✓ | ✓ | ✓ | ✓ |
| Alternative gas-based fuel technology 2 | ✗ | ✓ | ✓ | ✓ | ✓ | ✓ |
| CCS technology 1 | ✗ | ✗ | ✗ | ✓ | ✓ | ✓ |
| CCS technology 2 | ✗ | ✗ | ✓ | ✓ | ✓ | ✓ |
| EP-NETs technology 1 | ✗ | ✗ | ✗ | ✗ | ✓ | ✓ |
| EP-NETs technology 2 | ✗ | ✗ | ✗ | ✓ | ✓ | ✓ |
| EP-NETs technology 3 | ✗ | ✗ | ✓ | ✓ | ✓ | ✓ |
| EC-NETs technology 1 | ✗ | ✗ | ✗ | ✗ | ✓ | ✓ |
| EC-NETs technology 2 | ✗ | ✗ | ✗ | ✓ | ✓ | ✓ |
| EC-NETs technology 3 | ✗ | ✗ | ✓ | ✓ | ✓ | ✓ |

Based on Table 19, all renewable energies are available for deployment from Period 1 (except MSW which is absent in Period 1). Both technologies of alternative solid and gas-based fuels are available for deployment in Scenario 2, with their availability shown in Table 19. A similar situation also applied to both CCS technologies. Note that all NETs are now available for deployment, unlike Scenario 1. Technology 3 of both NETs having the highest $CO_2$ intensities versus technology 1 and 2 are available for deployment from Period 3 onwards. Meanwhile, technology 1 has the lowest $CO_2$ intensity (most effective for CDR) and thus is more mature compared to technology 2 and 3 but is only available in the final two periods.

### 5.2.1 Case 1

The information in Table 19 is inputted in the **'TECH_IMPLEMENTATION_TIME'** tab in the Microsoft Excel-based user interface file titled **'Base_User_Interface'**. Once again, the objective function of Case 1 is set to minimise the total $CO_2$ emissions (Equation 32). Figure 10 presents the results of Case 1. The detailed results of Case 1 in Scenario 2 are presented in Table S2 in the Supplementary Information.

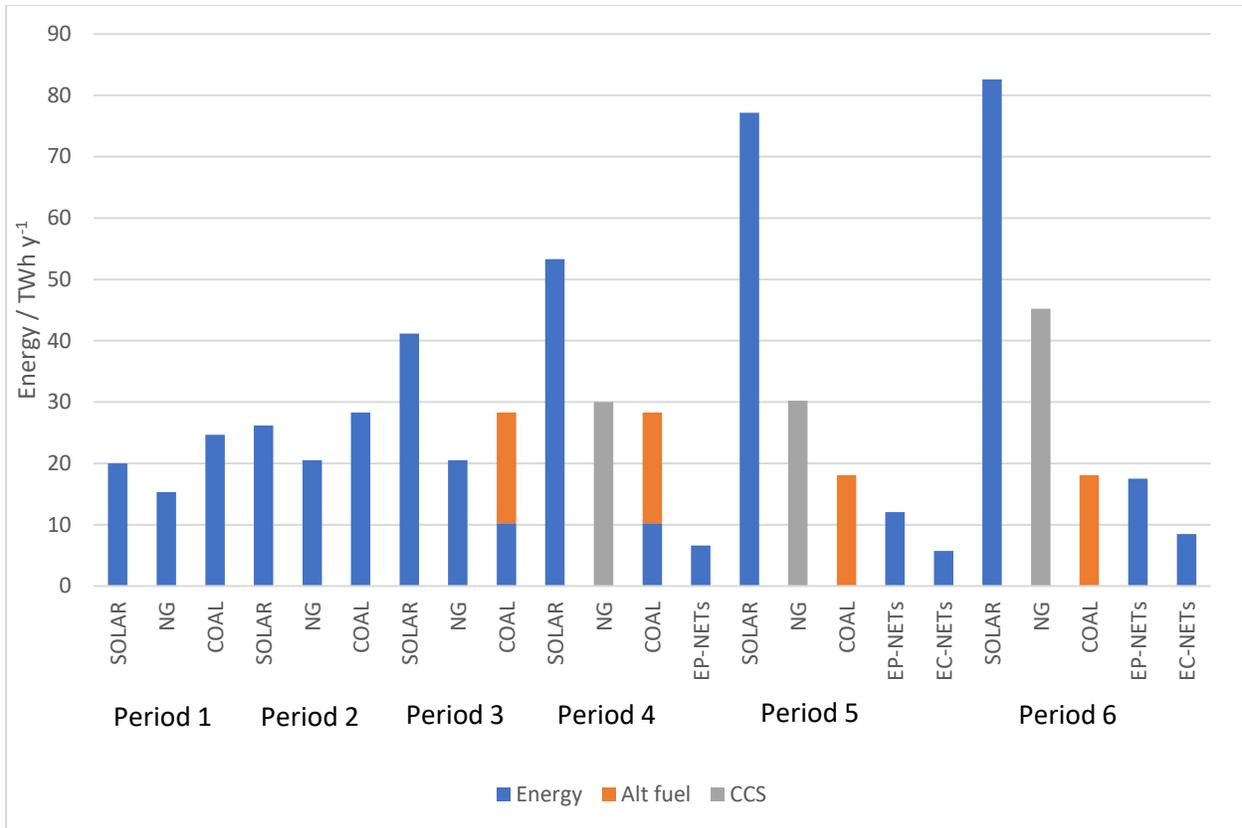

*Figure 10: Power plants configurations of Case 1 in Scenario 2*

Based on Figure 10, the configurations of the power plants in Periods 1 and 2 were identical to those observed in Scenario 1. Therefore, the total $CO_2$ load in both Periods 1 and 2 were identical to Scenario 1, thus violating the $CO_2$ emissions limits. Moving on to Period 3, both technologies of alternative fuels and CCS technology 2 were available for deployment. Therefore, coal in power plant 8 was replaced with a lower-carbon alternative. In Scenario 1, the fuels in power plants 3 and 8 were both substituted with alternative fuel technology 2 which is dirtier but cheaper. The use of the more expensive but cleaner alternative solid-based fuel technology 1 in power plant 8 meant there was an insufficient budget for fuel substitution to take place in power plant 3. Consequently, the $CO_2$ emission limit of 15 Mt y$^{-1}$ in Period 3 was violated.

Period 4 saw the availability of all mitigation technologies except for technology 1 concerning NETs. Like Scenario 1, CCS technology 2 was deployed for power plants 3 and 5, both fueled by natural gas. Additionally, 6.6 TWh y$^{-1}$ of EP-NETs technology 2 was deployed to aid in CDR. Nevertheless, the total $CO_2$ load may only be minimised to 21 Mt y$^{-1}$ which is 10 Mt y$^{-1}$ higher than the permissible limit. All mitigation technologies are available in the final two periods. The configurations of the power plants in Periods 5 and 6 were identical to those observed in Scenario 1. Aside from the deployment of CCS technology 2 and alternative solid-based fuel technology 1, EP-NETs technologies 1 and 2 and EC-NETs technology 1 were deployed. Technology 1 of both NETs was deployed due to their lower $CO_2$ intensities, thus contributing to a greater CDR. Note that the total $CO_2$ load in Period 5 was minimised at 5.9 Mt y$^{-1}$. This is the first period that saw the total $CO_2$ load below the $CO_2$ emissions limit of 6 Mt y$^{-1}$. Meanwhile, the net-zero emissions target was achieved in the final energy planning period. These results demonstrated that NETs deployment is crucial to achieving relevant climate change targets.

### 5.2.2 Case 2

Scenario 2 was repeated for Case 2 by minimising the total energy planning cost (Equation 31) as the objective function. The presence of a greater pool of mitigation technologies made it possible to solve Case 2 to global optimality. Note that ***'min_budget'*** is selected in cell 'B30' of the user-interface file (see Figure 6). Figure 11 presents the results of Case 2. The detailed results of Case 2 in Scenario 2 are presented in Table S3 in the Supplementary Information.

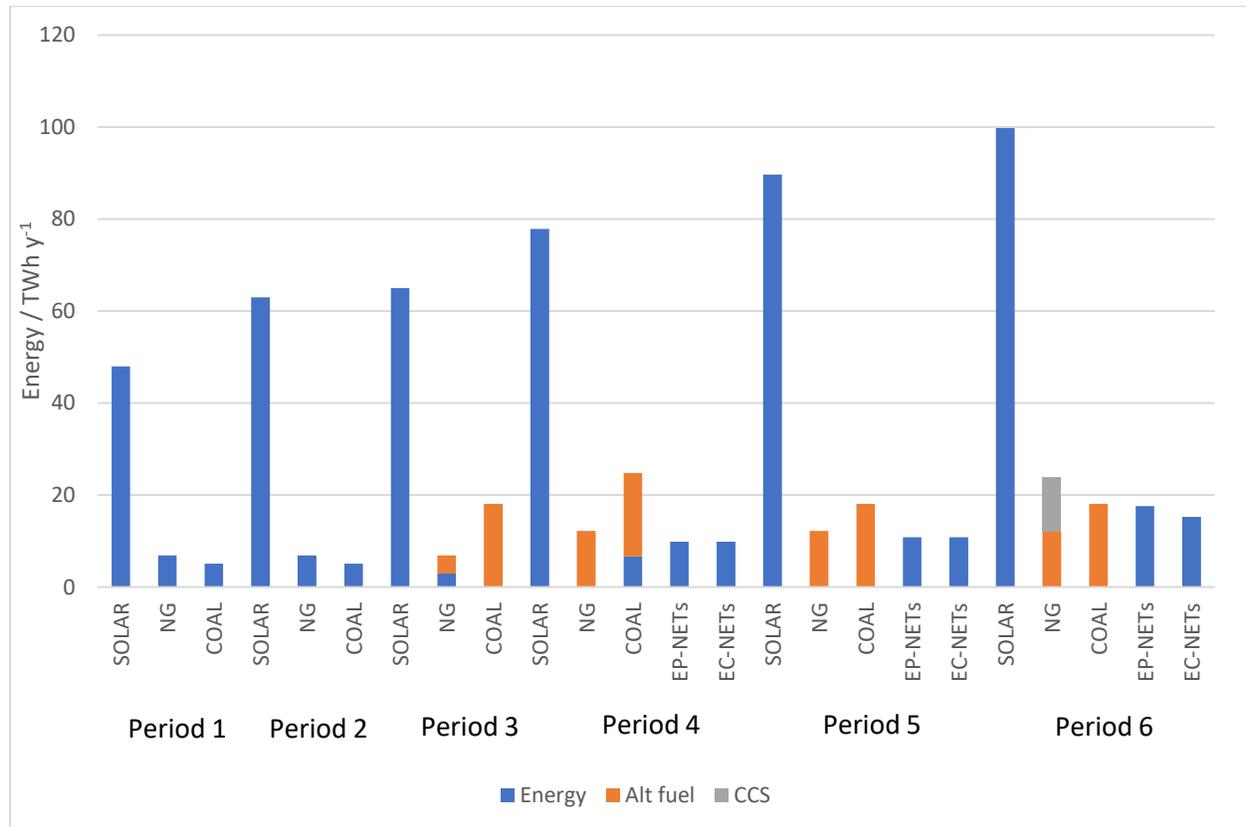

*Figure 11: Power plants configurations of Case 2 in Scenario 2*

Based on Figure 11, the $CO_2$ emissions limits for all periods were satisfied. In Period 1, power plants 1 and 2 (both fueled by solar), power plant 5 (natural gas) and power plant 8 (coal) were operational. Unlike Case 1, the deployment of both renewable plants resulted in the total $CO_2$ load being slightly below the $CO_2$ emission limit, thus satisfying the constraint. Note however that the total cost amounted to USD 3,673 mil $y^{-1}$, which is much higher than the allocated budget (USD 3,000 mil $y^{-1}$). These results explain the violation $CO_2$ emissions limit in Period 1 in Case 1. A greater budget allocation is required to satisfy the emissions constraint in the latter case. Moving on to Period 2, power plant 9 fueled by solar was commissioned to meet the demand of 75 TWh $y^{-1}$ and thus satisfy the $CO_2$ emissions limit of 18 Mt $y^{-1}$. Period 3 saw those fuels in power plants 5 (natural gas) and 8 (coal) being substituted for a lower carbon alternative. For both plants, technology 1 was deployed. Note that co-firing of the alternative gas-based fuel and natural gas occurred in power plant 5.

Period 4 saw the deployment of NETs, combined with greater power generation from power plant 5. Instead of co-firing that occurred in Period 4, the natural gas in power plant 5 was completely substituted with the alternative gas-based fuel technology 1. Note that technology 2 with lower $CO_2$ intensities were deployed for both types of NETs, despite being costlier. Since coal-based power plant 7 was decommissioned in Period 5, solar-fueled power plant 10 was deployed to meet the demand of 120 TWh y$^{-1}$. Period 6 saw the deployment of CCS technology 2 in power plant 4, fueled by natural gas as well as NETs technology 1. Once again, these results highlight the criticality of NETs deployment in mitigating $CO_2$ emissions.

## 6. Discussions

The two scenarios have demonstrated the usefulness of the *DECO2* software in performing optimal decarbonisation for the power generation sector. The software can consider the various commissioning and decommissioning timelines among power plants. Additionally, the *DECO2* software can help drive and inform decarbonisation strategy that can consider fuel substitutions and the deployment of CCS as well as NETs. Although the problems demonstrated in this paper are small-scale problems only involving 10 power plants, the problem may very easily be scaled for optimal decarbonisation pathways to be conducted on a national scale. Both scenarios were solved to global optimality in negligible time, thus promoting the use of *DECO2* in the industry. Additionally, the *DECO2* software has highlighted the importance of CCS and NETs deployment towards achieving the net-zero emissions targets and provides decision-makers with a free-to-use, easily modifiable tool. Most importantly, no single mitigation technology would be sufficient in achieving the relevant climate change targets, and hence a portfolio optimisation approach is required.

## 7. Conclusions

The *DECO2* optimal decarbonisation software framework was developed and introduced in this work to aid in carbon-constrained energy planning and the mitigation of $CO_2$ emissions. Consisting of a pool of available mitigation technologies such as alternative low-carbon fuels, renewable energies, CCS and NETs, the multiperiod energy planning model may be employed by policymakers and energy planners to determine the optimal deployment of each technology to meet the increasing power demand and stringent $CO_2$ emissions limits. The superstructural model in this work was developed in Python with an integrated user interface in Microsoft Excel. All energy planning data are inputted in the latter as the former only serves as optimisation software, meaning that the model is simple to use. The open-source framework allows flexibility to advanced users to change the formulation and input their constraints. Two scenarios with different availabilities of mitigation technologies are investigated in this work to demonstrate the software's functionality. The first scenario, which is the least aggressive approach, is optimised based on the minimised emissions for each period. Note that none of the periods satisfied the $CO_2$ emissions limit due to an insufficient budget. Meanwhile, Scenario 2 is considered more aggressive due to the greater availability of mitigation technologies. For Case 1 in Scenario 2, the $CO_2$ emissions were violated in the earlier periods, before achieving the net-zero target in the final period. Meanwhile, the results of Case 2 in Scenario 2 demonstrated that early deployment of renewable energy is crucial to ensure that mitigations of $CO_2$ emissions at later periods can be performed with relative ease. Once NETs are available, their deployment is crucial to aid in CDR. Future work should focus on the demand variation on a small scale i.e., daily, hourly etc. The superstructural model developed in this work has the potential to deal with demand peaks within a small timeframe. Also, practical development in the associated mitigation technologies must be integrated into the existing model to build a realistic energy planning scenario.

**Acknowledgement**

This work is partly a result of The British Council Japan's COP26 Trilateral Research Initiative entitled "*A Software Framework for Optimal Decarbonisation Planning for ASEAN Countries*" (https://www.britishcouncil.jp/en/programmes/higher-education/university-industry-partnership/cop26-trilateral-research-initiative) and the authors gratefully acknowledge the funding.